\documentclass[journal,draftclsnofoot,onecolumn,12pt]{IEEEtran}

\IEEEoverridecommandlockouts
\usepackage{cite}
\usepackage{amsmath,amssymb,amsfonts}

\usepackage{algorithm}
\usepackage{algorithmic}
\usepackage{graphicx}
\usepackage{textcomp}
\usepackage{xcolor}
\usepackage{geometry}
\usepackage{float}
\usepackage{lineno}
\usepackage{subfigure}
\usepackage{multirow}
\usepackage{array}

\usepackage{epsfig}
\usepackage{graphicx}

\usepackage{adjustbox}
\usepackage{array}
\usepackage{booktabs}
\usepackage{colortbl}
\usepackage{float,wrapfig}
\usepackage{hhline}
\usepackage{multirow}

\usepackage{amsmath,amsfonts,amsthm,amssymb}
\usepackage{bm}
\usepackage{nicefrac}
\usepackage{microtype}

\usepackage{changepage}
\usepackage{extramarks}
\usepackage{fancyhdr}
\usepackage{lastpage}
\usepackage{setspace}
\usepackage{soul}
\usepackage{xspace}

\begin{document}

\title{Symbol Rate and Carries Estimation in OFDM Framework: A high Accuracy Technique under Low SNR}


\author{
Zetian Qin, Yubai Li, Benye Niu, Qingyao Li and Renhao Xue
\vspace{-4mm}
}

\maketitle

\begin{abstract}
Under low Signal-to-Noise Ratio (SNR), the Orthogonal Frequency-Division Multiplexing (OFDM) signal symbol rate is limited. Existing carrier number estimation algorithms lack adequate methods to deal with low SNR. This paper proposes an algorithm with a low error rate under low SNR by correlating the signal and applying an Fast Fourier Transform (FFT) operation. By improving existing algorithms, we improve the performance of the OFDM carrier count algorithm. The performance of the OFDM's useful symbol time estimation algorithm is improved by estimating the number of carriers and symbol rate.

\end{abstract}


\section{Introduction}
In the next generation network (6G), signal processing technologies will be further developed \cite{9785899,zhao2022wknn,liu2022joint,wang2015new}. Among them, orthogonal frequency division multiplexing (OFDM) is a highlighted signal transmission technology \cite{hamamreh2020orthogonal}. 
OFDM divides channels into non-interference sub-channels, which can provide transmission according to their unique characteristics. Therefore, OFDM has an irreplaceable advantage in providing reliable communication \cite{9779965}, in the case of extreme low Signal-to-Interference plus Noise Ratio (SINR) \cite{8792710}. Therefore, OFDM technology is widely used in dense cellular networks \cite{lou2021green} and ultra-long distance transmission \cite{liu2022machine,wang2022ultra}.\par
For non cooperative communication \cite{ke2019blind}, there are three steps to process when receiving unknown signals. (\romannumeral1) Determine the modulation mode of unknown received signal. If the modulation mode of the received signal is OFDM modulation, we should carry out the parameter blind estimation. (\romannumeral2) Extract the characteristic quantities of the received signal and use these characteristic quantities to estimate the carrier frequency, symbol rate, symbol time and other parameters. (\romannumeral3) Decode the unknown received signal by using the parameters obtained by the estimation method and restore the original symbols. At present, the technology of modulation recognition is relatively mature \cite{azzouz2013automatic,nandi1998algorithms,azzouz1996modulation,o2016convolutional,chen2019toward}. In engineering, the technology of restoring signal symbols is very complete \cite{selesnick2009signal,rajesh2020channel,}.
However, one of the challenges of OFDM is how to extract the required parameters from a blind signal. Because of its complex modulation mode, a lot of key parameters need to be estimated \cite{5725488,guo2020securing,wang2017constellation}, such as the symbol rate, useful symbol time, and the number of carries.

\par

For the above key parameters, existing literature has proposed estimation methods. For useful symbol time, \cite{1544022,6240069,4411681} provided different estimation methods, respectively: (\romannumeral1) \cite{1544022} is based on the auto-correlation feature of the OFDM signal's Cyclic Prefix (CP), (\romannumeral2) the authors in \cite{6240069} used high-order cumulants combined with wavelet transform, (\romannumeral3) the authors in \cite{4411681} obtained the CP length and useful symbol time through maximum likelihood. Spectrum symmetry algorithm in \cite{4411681} 
and cyclic spectrum algorithm in \cite{6240069} are used to obtain the carrier frequency. The second-order cyclic cumulant is used to derive the the number of carriers \cite{4221485}. However, the performance of high-order cumulants is poor at low Signal-to-Noise Ratio (SNR). Although the above methods have provided a mature scheme for parameter estimation of OFDM, their performance will not be reliable when dealing with low SNR \cite{1544022,6240069,4411681,4221485}. To overcome this challenge, this paper analyzes the relationship between number of carries of OFDM signal spectrum and data length. Under low SNR, multi segment data superposition is used to reduce the influence of noise in the power spectrum and design a traversal algorithm to estimate the total symbol time of OFDM signal.For the oversampled signal, this use traditional method to estimate the oversampling rate and design a alternative algorithm to estimate the number of carries of OFDM signal. Compared with the traditional method, the substitution method and traversal method designed in this paper have better performance under low SNR. 

\begin{table*}[]
\begin{small}
\centering
\caption{Summary of Notations}
\begin{tabular}{c|c}

\hline

\textbf{Notation}                                            & \textbf{Description}        \\ \hline  \hline
$\sigma_{s}^{2}$; $N_{s}$ & Average power of signal; Total number of points of a single OFDM symbol.                               \\ \hline
 $N_{u}$; $N_{g}$ &  Points of a single useful OFDM symbol; Points of CP in a single OFDM symbol.                               \\ \hline
$N_{O}$; $N_{P}$ &  Amount of OFDM symbols; Parameter used to estimate $N_{s}$. \\ \hline

$N_{os}$ &  Amount of points of a symbol after oversampling. \\ \hline

$N_{ch}$; $N_{cn}$ &  Number of data segments with $N_{o}\cdot N_{P}$ data points; Number of carriers of received signal. \\ \hline

$i$; $k$ & Time domain abscissa; Starting point of time domain accumulation operation. \\ \hline
$j$; $i_{m}$ & Imaginary unit; Number of symbols of each piece of data in \eqref{eq4}. \\ \hline
$L_{M}$; $L'$; $L$ &  Points of using data; Length of the sliding window; Value set in advance between $N_{s}$ and $L_{M}$. \\ \hline
$k_{\rm{all}}$; $k_{\max}$ & Abscissa of all peaks in $Y(n,N_{P})$; Maximum value of $k_{\rm{all}} $. \\ \hline

 $N_{\rm{all}}$; $L_p$ & Amounts of $k_{\rm{all}}$; Peak spacing. \\ \hline
 $N_{\rm{use}}$; $L_{\rm{cur}}$ & Recorded peak; Temporary peak spacing. \\ \hline
 $k_{\rm{cur}}$; $N_{\rm{cur}}$ & Abscissa of current statistics; Amount of temporary counted peaks. \\ \hline
 $k_{i}$; $k_{j}$ &Abscissa of $k_{\rm{all}} $. \\ \hline
 $N_{\max}$; $N_{\min}$ &Possible maximum and minimum value of $N_{P}$. \\ \hline
 $P_u$; $N_{PL}$ &The array used to store useful $N_{P}$; Length of array $P_u$. \\ \hline
 $q$; $L_{p}$ &Oversampling rate; Distance between the peaks of $F_{r} \left ( k,\tau  \right )$. \\ \hline

\end{tabular} 
\end{small}
\end{table*}

\section{Calculation of symbol rate by traversal method}

Based on the traditional model, this paper makes some improvements. First, we introduce the traditional parameter estimation method. According to \cite{6100719}, the useful symbol length of OFDM is estimated by auto-correlation of the signal. The model is given as follows,

\begin{equation}
 \mathbb{E}\left \{ r\left ( i \right ) r^{\ast}\left ( i+k \right )  \right \} =\left\{
\begin{aligned}
    &\sigma_{s}^{2}, \ &k=0,  \\
     \sigma _{s}^{2} & \cdot \frac{N_{g}}{N_{s}} , \ &k=N_u, \\
    &0, \ &\mathrm{others,}
\end{aligned}\right.\label{eq1}
\end{equation}
where $\sigma _{s}^{2}=\mathbb{E}\left \{ \left | s(n) \right | ^{2}  \right \}$, $N_{g}$ are points of cyclic prefix (CP) in a single OFDM symbol, $*$ represents the conjugate of the signal, $k$ is the delay of $r\left ( i \right )$, $N_s$ is the number of points of a single OFDM symbol, and $N_u$ is the number of useful symbols. $N_u$ can be found by \eqref{eq2}. Note that the OFDM useful symbol time can be given as $N_u$ divided by the sampling rate. \eqref{eq1} shows that $\mathbb{E}\left \{ r\left ( i \right ) r^{\ast}\left ( i+k \right ) \right \}$ reaches its second largest value when $k$ is equal to $N_u$. Therefore, we traverse $k$ and find the second peak of $\mathbb{E}\left \{ r\left ( i \right ) r^{\ast}\left ( i+k \right )  \right \}$ by the following formula \cite{6100719}:
\begin{equation}
\begin{aligned}
    N_u=\mathop{\arg\max}\limits_{k}\left (  \frac{\left| \sum_{i=1}^{L_M-L'} r\left ( i \right ) r^{\ast } \left ( i+k \right ) \right| }{\frac{1}{2} \sum_{i=1}^{L_M-L'} \left (  \left | r\left ( i \right ) \right |^{2} + \left | r^{\ast } \left ( i+k \right ) \right | ^{2}\right )}  \right)
    ,\\  
    k=1,2,...,L,
\end{aligned}
    \label{eq2}
\end{equation}
where $L_{M}$ are points of using data, $L$ is a value set in advance between $N_{s}$ and $L_{M}$, the denominator represents the energy of the received signal, $L'$ is the length of the sliding window, and the molecule represents the sum of the auto-correlation of the received signal.

\par

For the estimation method of the symbol rate, we generally estimate the total OFDM symbol time, because its reciprocal is the symbol rate. The formula in \cite{6100719} is as follows,
\begin{equation}
\begin{split}
    L\left ( k \right )  =  \frac{\sum_{i=1}^{L'} r\left ( i+k \right ) r^{\ast }\left ( i+k+N_u \right ) }{\sum_{i=1}^{L'} \left( \left | r\left ( i+k \right )  \right |^{2}+\left | r\left ( i+k+N_u \right )  \right |^{2}\right )}  , \\
    k=1,2,...,L_M-N_u-L',\label{eq3}
\end{split}
\end{equation}
where $k$ indicates the location of the data point, $i$ represents the data position of the sliding window, $r\left ( i+k \right )$ represents the $i^{th}$ data of the sliding window.  $N_s$ is the distance between the midpoint of adjacent peaks of $L\left ( k \right )$. \par

\begin{figure}[H]
	\centering
    \includegraphics[width=0.7\linewidth]{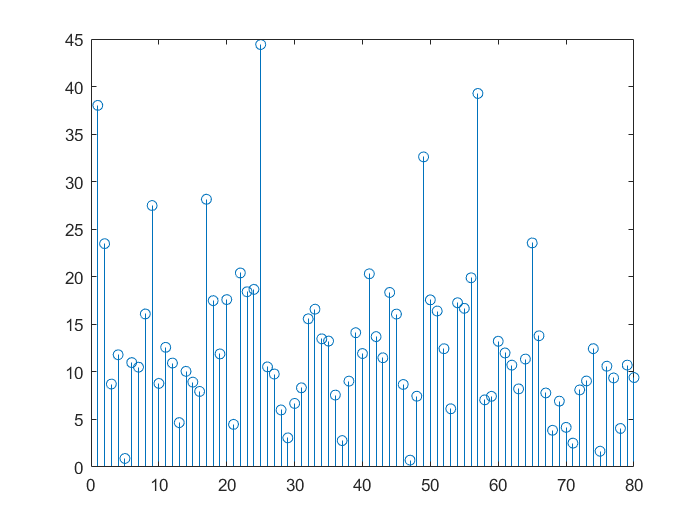}
	\caption{FFT of the received signal.}
	\label{fft}
\end{figure}

For OFDM signals, when the number of data points used is $N_s\cdot N_o$, where $N_o$ are the number of symbols, and the initial point of data used is the initial point of a single symbol, each sub-carrier has a corresponding peak on the frequency spectrum. The spacing between corresponding peaks $N_o$, which is shown in the Fig. \ref{fft}.\par

\begin{figure}[H]
	\centering
    \includegraphics[width=0.7\linewidth]{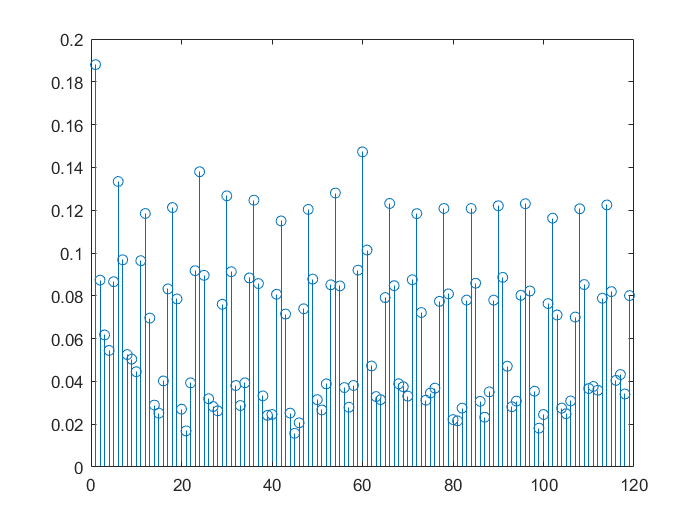}
	\caption{Convolution of frequency domain signal.}
	\label{juanji}
\end{figure}

 After performing auto-correlation operation on the data points of the received signal and then FFT operation, the result is as shown in Fig. \ref{juanji} .
In order to further reduce the influence of noise, we divide the data used into multiple segments. These data segments start with the initial point of a single OFDM symbol. The following formula can be obtained by summing each segment of data after auto-correlation operation and then FFT operation.
\begin{equation}
\begin{aligned}
    y\left ( i,N_{P} \right ) =\frac{1}{N_{ch}} \sum_{i_{m}=1}^{N_{ch}} {r\left ( N_{P}\cdot \left ( i_{m}-1 \right )+i   \right ) }  \\
  \times {r^{\ast } \left ( N_{P}\cdot \left ( i_{m}-1 \right )+i  \right )},\\
 i=1,2,... ,N_o \cdot N_{P},
\end{aligned}
    \label{eq4}
\end{equation}
\begin{equation}
    Y\left(n,N_{P}\right)=\left | \sum_{i=1}^{L_M} y(k,N_{P})e^{-jni} \right |  , \label{eq5}
\end{equation}
where $N_{ch}$ is the number of chips, $N_o$ is number of chips observed. When $N_{P}$ is the number of points of a single symbol, the peak whose abscissa is an arithmetic progression will appear on $Y\left ( n,N_{P} \right )$. At low SNR, the peaks still exist, while the amplitudes of peaks become lower. When $N_{P}$ is not equal to $N_{s}$, peaks that belong to arithmetic progression account for a reduced proportion of all peaks.
\begin{figure}[H]
	\centering
\includegraphics[width=0.7\linewidth]{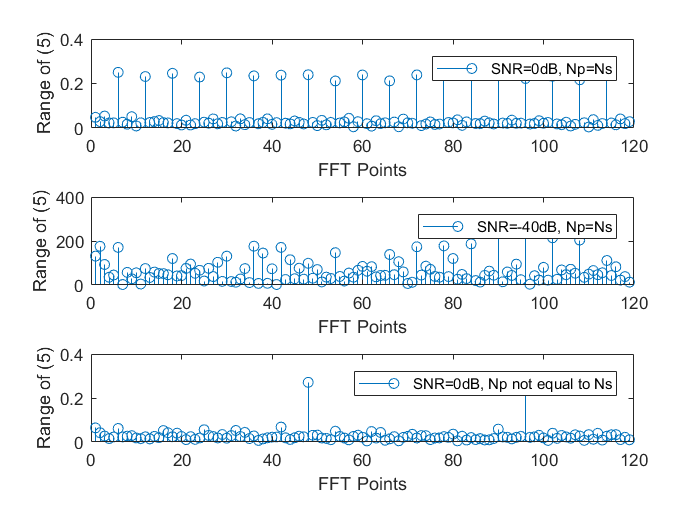}
	\caption{Comparison diagram under different SNR and $N_{P}$.}
	\label{FFT image}
\end{figure}

Then we design a method to find $N_s$. According to \eqref{eq4} and \eqref{eq5}, the traversal algorithm is proposed: (\romannumeral1) traverse $N_{P}$, (\romannumeral2) find the abscissa of the peak of $Y\left( n,N_{P} \right)$, (\romannumeral3)  Use algorithm \ref{alg:1} to find out the number of peaks with fixed spacing and the total number of peaks, (\romannumeral4) use algorithm \ref{suanfa2} to find $N_{P}$ which satisfy the judgment conditions and (\romannumeral5) calculate the mean value of all $N_{P}$ that meet the judgment conditions, and judge the calculation result as $N_{s}$. The specific algorithm processes are shown in algorithm \uppercase\expandafter{\romannumeral1} and algorithm \ref{suanfa2}.

\begin{algorithm}[t]
\begin{algorithmic}\label{alg:1}
\caption{$\left [  N_{\rm{use}},N_{\rm{all}},L_{p}\right ] $ = functionA $\left (  Y\left ( n,N_{P} \right ) \right ) $}
    \STATE \textbf{(A) Initialize:} Record the abscissa of all peaks in $Y(n,N_{P})$ (denoted as $k_{\rm{all}}$), the maximum value of $k_{\rm{all}} $ (denoted as $k_{\max}$), amounts of $k_{\rm{all}}$  array $N_{\rm{all}}$, peak spacing $L_p=0$, recorded peak $N_{\rm{use}}=0$, temporary peak spacing $L_{\rm{cur}}=0$, the abscissa of current statistics $k_{\rm{cur}}=0$, amount of temporary counted peaks $N_{\rm{cur}}=0$.
    \STATE \textbf{(B) Find the number and spacing of the peaks whose abscissa satisfies the arithmetic sequence:}

    \FOR{$k_{i}=1;k_{i}<\frac{N_{\rm{all}}}{2};k_{i}=k_{i}+1$}
    \FOR{$k_{j}=k_{i}+1;k_{j}<\frac{N_{\rm{all}}}{2}+2;k_{j}=k_{j}+1$}
    \IF{$k_{\rm{all}}(k_{j})-k_{\rm{all}}(k_{i})>3$}
    
    \STATE $N_{\rm{cur}}=2$.
    \STATE $L_{\rm{cur}}=k_{\rm{all}}(k_{j})-k_{\rm{all}}(k_{i})$. 
    \STATE $k_{\rm{cur}}=k_{\rm{all}}(k_{j})$.
    
    \WHILE{$k_{\rm{cur}}<k_{\max}$}
    \STATE $k_{\rm{cur}}=k_{\rm{cur}}+L_{\rm{cur}}$.
    \IF{$L_{\rm{cur}}$ is a member of $k_{\rm{all}}$}
    \STATE $N_{\rm{use}}=N_{\rm{use}}+1$.
    \IF{$N_{\rm{cur}}>N_{\rm{use}}$}
    \STATE $N_{\rm{use}}=N_{\rm{cur}}$.
    \STATE $L_p=L_{\rm{cur}}$.
    \ENDIF
    \ENDIF
    \ENDWHILE
    \ENDIF
    \ENDFOR
    \ENDFOR
    \RETURN $\left [  N_{\rm{use}},N_{\rm{all}},L_{p}\right ] $.
    
\end{algorithmic}
    
\end{algorithm}


\begin{algorithm}[t]
\begin{algorithmic}\label{suanfa2}
\caption{Traversal method}

    \STATE \textbf{(A) Initialize:} Initialize value $N_{P}$ and its possible minimum and maximum value $N_{\min}$ and $N_{\max}$. Record the array used to store useful $N_{P}$ (Denoted as $P_u$). Let $N_{P}=N_{\min}$, $N_o=6$.
    \STATE \textbf{(B) Calculate the distance and number between the peaks of $Y(n,N_{P})$ under the current $N_{P}$:}
    \STATE Generate $Y\left (n,N_{P}\right )$.
    \STATE $\left [  N_{\rm{use}}, N_{\rm{all}},L_{p}\right ] $ = functionA $\left (  Y\left ( n,N_{P} \right ) \right ) $.

    \STATE \textbf{(C) Determine whether the data is useful:} 
    \IF{$\frac{N_{\rm{use}}}{N_{\rm{all}}} >0.5$ and $N_{all}>N_{min}$ and $L_p=N_o$}
    \STATE Store $N_{P}$ in $P_u$, $N_{P}=N_{P}+1$.
    \IF{$N_{P}<N_{\max}$}
    \STATE Regenerate $Y\left(n,N_{P} \right)$ .
    \STATE $\left [  N_{\rm{use}},N_{\rm{all}},L_{p}\right ] $ = functionA $\left (  Y\left ( n,N_{P} \right ) \right ) $.
    \STATE Go to step (C).
    \ENDIF

    \ELSE
    \IF{There are continuous values in $P_u$ tail}
    \STATE Remove discontinuous values in $P_u$.
    \STATE Go to step (D).
    \ELSE
    \STATE $N_{P}=N_{P}+1$.
    \IF{$N_{P}<N_{\max}$}
    \STATE  Regenerate $Y(n,N_{P})$ and $k_{\rm{all}}$.
    \STATE $\left [  N_{\rm{use}},N_{\rm{all}},L_{p}\right ] $ = functionA $\left (  Y\left ( n,N_{P} \right ) \right ) $.
    \STATE Go to step (C).
    \ENDIF
    \ENDIF
    \ENDIF

    \STATE \textbf{(D) Determine the final value:} 
    \STATE $N_{PL}$ is the length of array $P_u$. 
    \RETURN $N_s=\sum_{k=1}^{N_{PL}}\frac{1}{N_{PL}}P_{u}\left(k \right)$.
\end{algorithmic}
    
\end{algorithm}

At last, the improved method achieves 75\% accuracy and its Amplitude Error (AE) is between 0.1 and 0.2 under low SNR.

\section{Calculation of number of carries and useful symbol time by substitution method}

\par
For the over-sampled signal, its data points need to be separated before using the above method to estimate $N_{s}$. According to \cite{6100719}, the number of carriers and the oversampling rate are estimated by oversampling the signal and performing a correlation operation, which can be expressed as follows,
\begin{equation}
    F_{r} \left ( \omega ,\tau  \right ) = \lim_{L_M \to \infty } \sum_{i=0}^{L_M} \mathbb{E} \left \{ r\left ( i \right ) r^{\ast}\left ( i+\tau  \right )    \right \} \exp\left ( -j\tfrac{2\pi }{L_M}\omega i   \right ) ,\label{eq6}
\end{equation}
where $L_M$ is the amount of using data, $j$ is imaginary part, $r\left ( i \right )$ is an over-sampled signal, $F_{r}$ represents power spectral density. Generally, when $\tau=1$, the power spectral density is used in \eqref{eq7}. After the above analysis, the estimation formula of oversampling rate is as follows,
\begin{equation}
    q=\frac{L_M}{L_p},\label{eq7}
\end{equation}
where $L_p$ is the distance between the peaks of $F_{r} \left ( k,\tau  \right )$. After calculating $q$, the signal's data points are separated to restore the original signal. The symbol length $N_{os}$ of the original signal can be obtained by using the algorithm \ref{alg:1} to process the OFDM received signal without over sampling. The estimation result formula of OFDM carrier number is as follows,
\begin{equation}
    N_{cn}=2^ {\left ( \left \lfloor \log_{2}{\left ( N_{os} \right )  } \right \rfloor  \right )},  \label{eq8}
\end{equation}
where $N_{cn}$ is the OFDM carrier number, $\left \lfloor \cdot \right \rfloor$ represents rounding down to an integer, $q$ is the oversampling rate. After calculating the number of carriers, the length of the OFDM useful symbol can be obtained as follows,
\begin{equation}
    N_{u}=q\cdot N_{cn}.  \label{eq9}
\end{equation}
\begin{figure}[H]
\centering
	\includegraphics[width=0.7\linewidth]{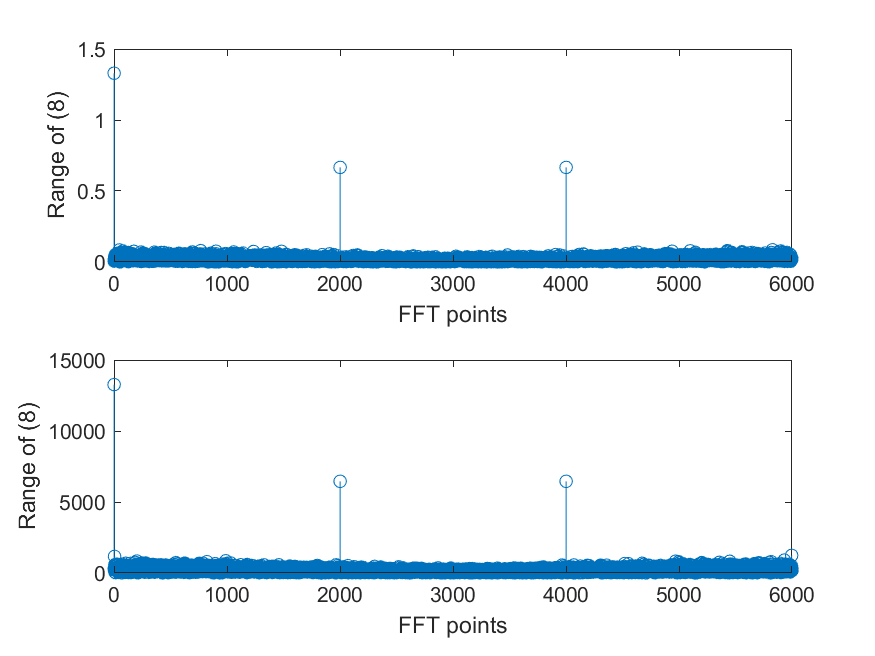}
\caption{FFT of correlation signal after oversampling when the SNR(dB) is 0 and -40.}
\label{fig}
\end{figure}
\par
The image calculated by \eqref{eq6} is less affected under low SNR. Use $N_s$ by algorithm \ref{alg:1} under low SNR can improve the accuracy of the algorithm and reduce the error.

\begin{algorithm}[t]
\begin{algorithmic}
\caption{Substitution method}
    \STATE \textbf{(A) Initialize:}
    \STATE Input received signal $r \left(i \right)$.
    \STATE \textbf{(B) Generate $F_{r}$ and calculate $q$:} 
    \STATE $F_{r} \left ( k,\tau  \right ) =\lim_{M \to \infty } \sum_{n=0}^{M} \mathbb{E} \left \{ r\left ( i \right )\cdot r^{\ast}\left ( i+\tau  \right )    \right \}\cdot$
    
    $\exp\left ( -j\tfrac{2\pi }{M}kn   \right )$ , $q=\frac{L_M}{L_P}$.
    \STATE \textbf{(C) Calculate $N_{os}$.}
    \STATE Under-sample $r \left(i \right)$ and use the result by \ref{alg:1} to calculate $N_{os}$.
    \STATE \textbf{(D) Determine the number of carriers:} \\ 
     $N_{cn}=2^ {\left ( \left \lfloor \log_{2}{\left ( N_{os} \right )  } \right \rfloor  \right )}$.
     \STATE \textbf{(E) Determine $N_u$ according to \eqref{eq8}:} \\ 
     \RETURN $N_{u}=q\cdot N_{cn} $.
    
    \end{algorithmic}
    \label{suanfa3}
\end{algorithm}

\section{Experiment and results}

\begin{figure}[H]
\centering
	\includegraphics[width=0.7\linewidth]{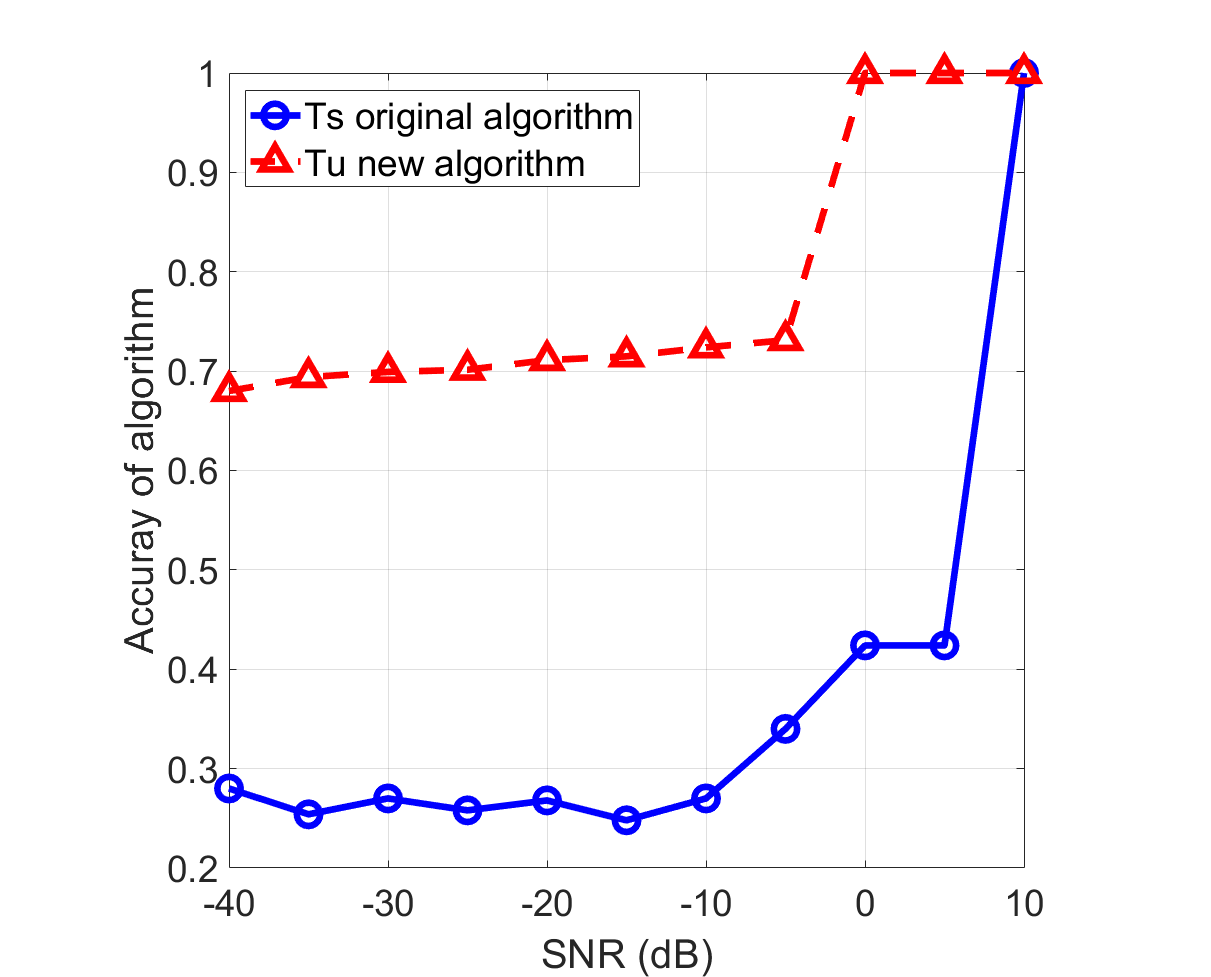}
    \caption{Comparison of symbol length algorithm accuracy.}
\label{fig4}
\end{figure}

In this experiment, the modulation mode is 16 Quadrature amplitude modulation (QAM). The number of carriers is 128, the proportion of CP is 0.25. The carrier frequency is 10M. The sampling rate is 40M. The number of symbols is 20. The channel is the Gaussian channel. Set the SNR from -40dB to 10dB. The number of tests under each SNR is $10^5$. When SNR is lower than -5dB, we choose \ref{alg:1} to calculate the $N_s$. When SNR is higher than -5dB,  we choose \eqref{eq3} to calculate the $N_s$.

\begin{figure}[H]
\centering
	\includegraphics[width=0.7\linewidth]{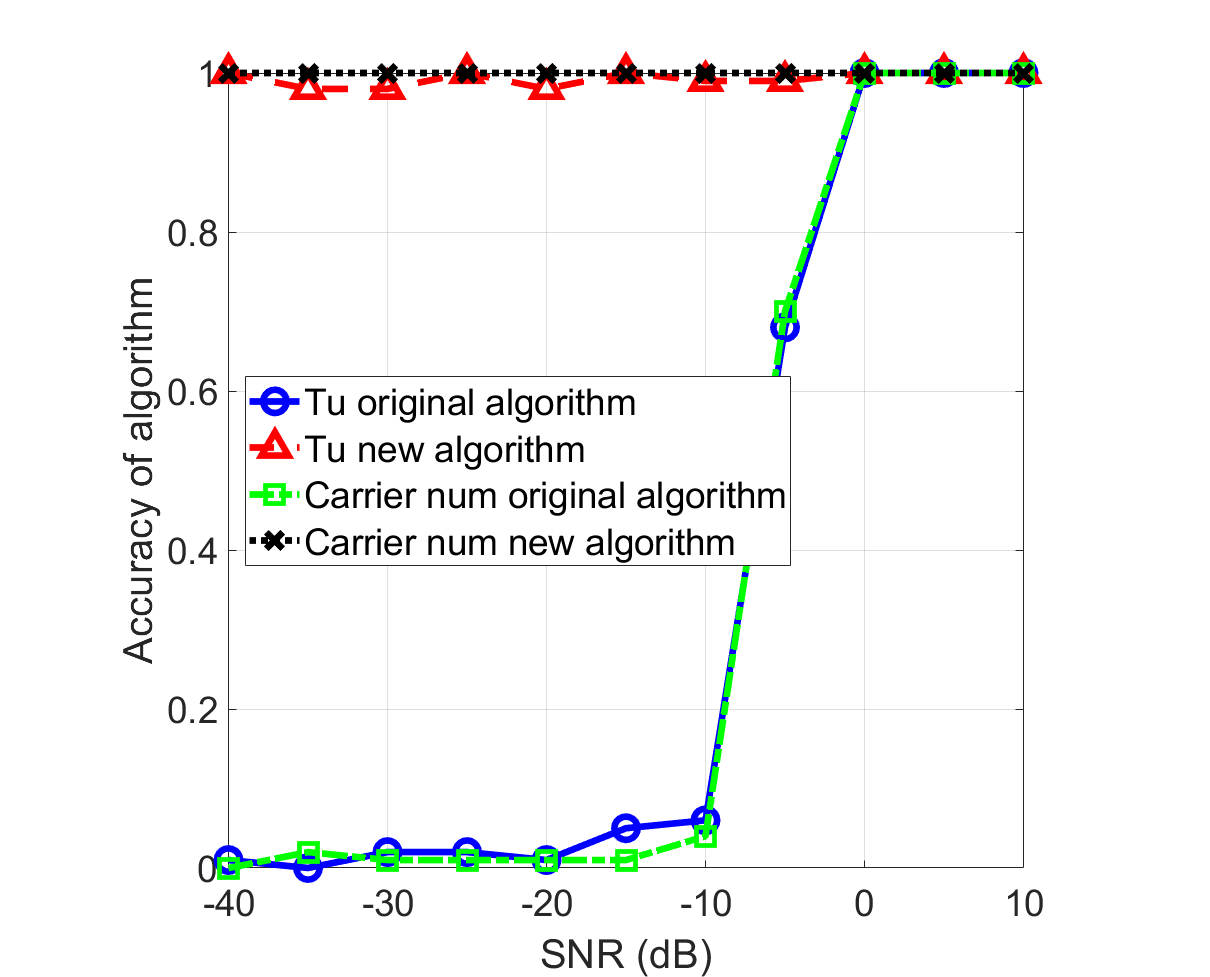}
\caption{Comparison of useful symbol length and carrier number algorithm accuracy.}
\label{fig5}
\end{figure}

By observing the results of the experiments, it can be found that under high SNR, the accuracy of $N_{s}$ and $N_{u}$ are not as good as \eqref{eq2} and \eqref{eq3}, and their AEs are about 0.1. The AEs of two algorithms of carrier number have the same performance. At low SNR, the accuracy of the new estimation method for $N_{s}$ and $N_{u}$ is between 0.9 and 1, and the AE is between 0 and 0.15. Compared with the original algorithm, the performance is significantly improved. The accuracy of the carrier number can reach 1 and the AE is 0.\par

\begin{figure}[H]
\centering
	\includegraphics[width=0.7\linewidth]{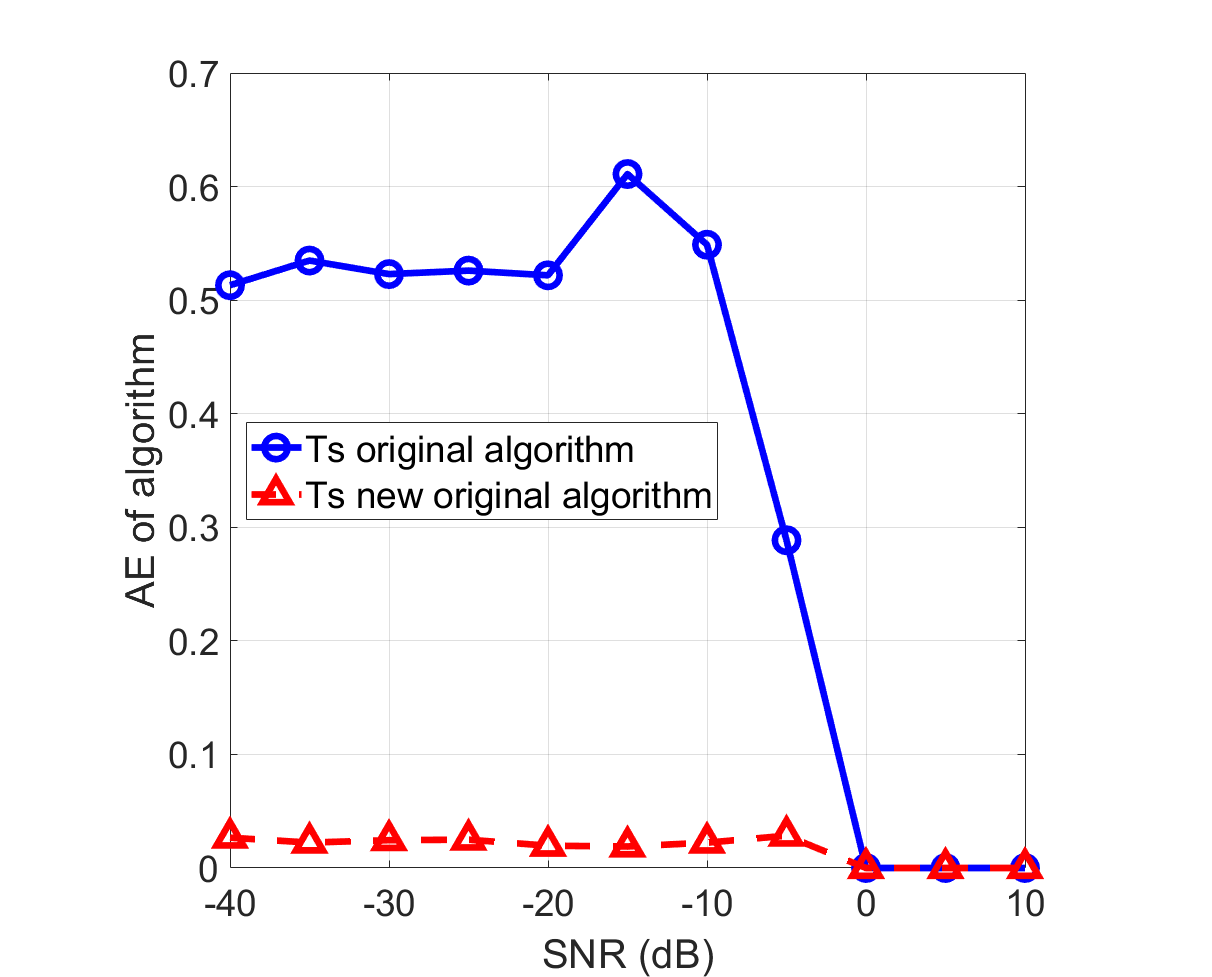}
\caption{Comparison diagram of symbol length algorithm AE.}
\label{fig6}
\end{figure}

The accuracy and AE of $N_s$ fluctuate under low SNR. In the experiment, $N_s$ is set to 160. Members in $P_{u}$ include $N_{s}$, but their mean value may not be equal to ns. In this case, it is challenging to determine which value is the accurate number of OFDM symbol points. The proposed algorithm improves the AE performance under low SNR.

\begin{figure}[H]
\centering
	\includegraphics[width=0.7\linewidth]{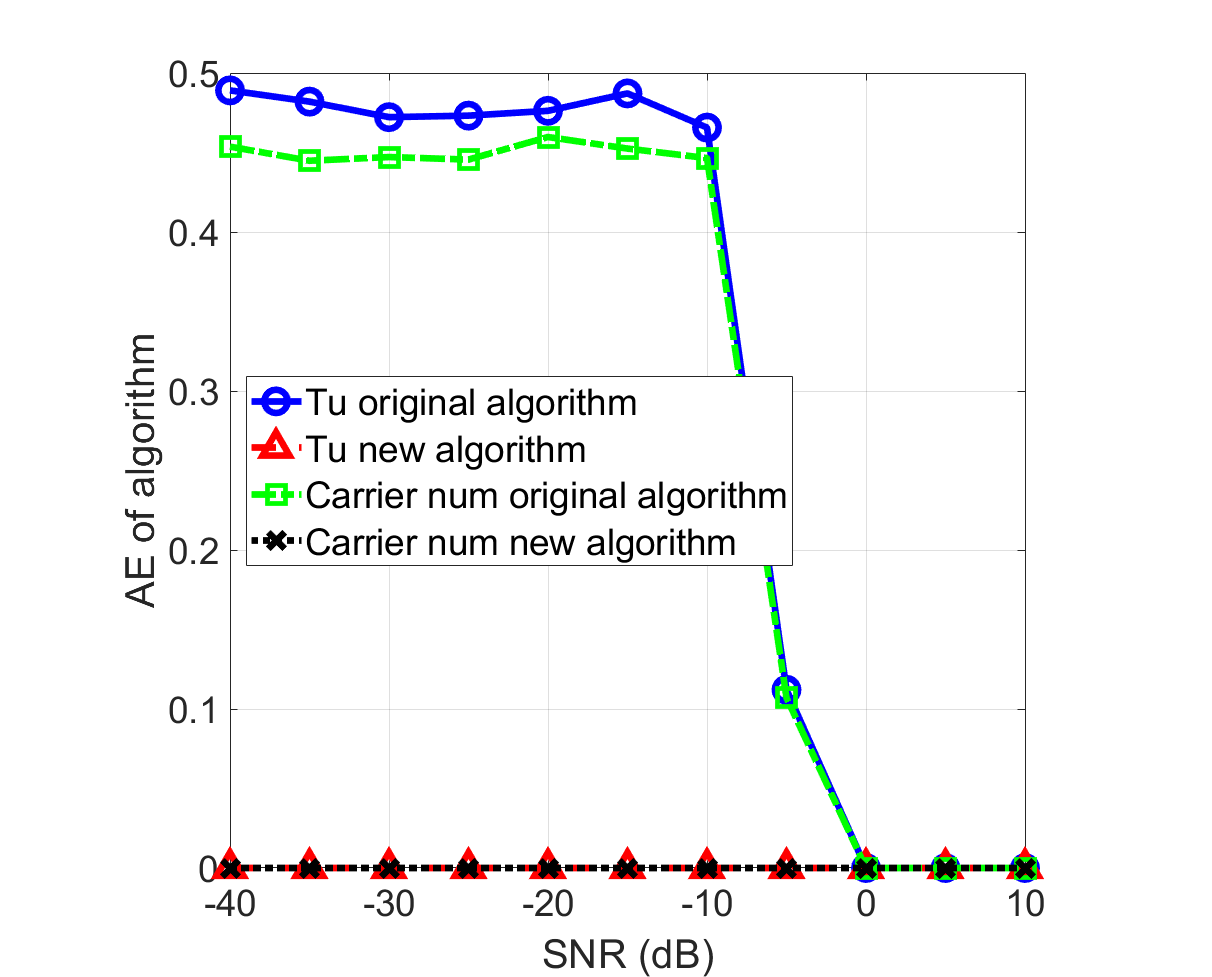}
\caption{Comparison diagram of symbol length and carrier number algorithm AE.}
\label{fig7}
\end{figure}

\section{Conclusion}
This paper proposed an improved algorithm for the total OFDM symbol time $N_s$ (reciprocal of symbol rate). Under low SNR, the estimation accuracy has been improved. However, the OFDM useful symbol time that used in the the oversampling rate algorithm is greatly affected by noise. In order to avoid the situation above and improve the performance, the useful symbol time is replaced with the total symbol time. OFDM signal has the characteristic that the number of carriers is exponential of 2. The proposed algorithm can achieve high accuracy. After estimating the number of carriers, the OFDM useful symbol time can be deduced.
\bibliographystyle{IEEEtran}
\bibliography{references.bib}

\end{document}